\documentclass[twocolumn,aps,pre,amsmath,amssymb,amsfonts,amstex,floatfix,superscriptaddress]{revtex4}
\usepackage{graphicx}
\usepackage{dcolumn}
\usepackage{bm}
\usepackage[hypertex]{hyperref}
\usepackage{latexsym}
\usepackage{float}
\usepackage{supertabular}
\usepackage{longtable}

\begin{document} 

\title{First-Passage Properties of  Bursty Random Walks}
\title{J. Stat.\ Mech.\ P06018  (2010)}
\author{D. Volovik}
\affiliation{Center for Polymer Studies and Department of Physics, Boston University, Boston,
Massachusetts 02215 USA}
\author{S. Redner}
\affiliation{Center for Polymer Studies and Department of Physics, Boston University, Boston,
Massachusetts 02215 USA}

\begin{abstract}

  We investigate the first-passage properties of bursty random walks on a
  finite one-dimensional interval of length $L$, in which unit-length steps
  to the left occur with probability close to one, while steps of length $b$
  to the right --- ``bursts'' --- occur with small probability.  This
  stochastic process provides a crude description of the early stages of
  virus spread in an organism after exposure.  The interesting regime arises
  when $b/L\alt 1$, where the conditional exit time to reach $L$,
  corresponding to an infected state, has a non-monotonic dependence on
  initial position.  Both the exit probability and the infection time exhibit
  complex dependences on the initial condition due to the interplay between
  the burst length and interval length.
\end{abstract}

\maketitle

\section{Introduction}

We are continually exposed to viruses.  Despite these constant biological
assaults, the immune system successfully fends off most viruses.
Considerable effort has been devoted to modeling the factors that influence
whether a person exposed to a particular virus will eventually become
ill~\cite{nowak}.  Typical theoretical models of viral infections account for
the evolution of the number of infected cells, healthy cells, and viruses as
a function of the rates of microscopic infection and transmission rates.
Such models have provided many useful insights about the dynamics of viral
diseases~\cite{nowak,perelson}.

\begin{figure}[ht]
  \centerline{\includegraphics*[width=.4\textwidth]{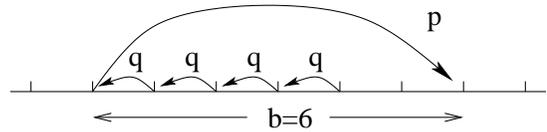}}
  \caption{A bursty random walk with burst length $b=6$.}
  \label{model}
\end{figure}

In this work, we study a toy model --- the bursty random walk
(Fig.~\ref{model}) --- that captures one of the elements of viral infection
dynamics.  The position of the walk in one dimension represents the number of
active viruses in an organism.  Since the immune system constantly kills
viruses, they are removed from the body at some specified rate, corresponding
to steps to the left in the bursty random walk.  However, with a small
probability, a virus enters and successfully hijacks a cell, the outcome of
which is a burst of a large number of new viruses into the host organism,
corresponding to a long step to the right in the model.

When the number of virus particles reaches zero, the organism may be viewed
as being free of the disease.  Conversely, when the number of viruses reaches
a threshold value $L$, the organism can be viewed as either being ill or
dead.  With this simplistic perspective, being cured or becoming ill is
recast as a first-passage problem for the bursty random walk in an interval
of length $L$.  When the burst length $b$ is small, the walk has a diffusive
continuum limit whose first-passage properties are well known~\cite{vk,fpp}.
However, if the burst length is of the order of the system length, this
burstiness effects strongly affect the first-passage characteristics.  This
large-burst limit should be applicable to infectious processes where the
threshold number of viruses for being ill is not large and the number of new
viruses created in a burst event is a finite fraction of this
threshold~\cite{PKP10}.  Related discreteness effects were found in the
first-passage characteristics of a random walk that hops uniformly within a
range $[-a,a]$ in the interval $[0,L]$, with $a\alt L$~\cite{AR06}.

In the next section, we define the model and the basic first-passage
quantities that we will investigate.  In Secs.~\ref{sec:E} and \ref{sec:t},
we determine the exit probabilities and the average exit times to either end
of the interval as a function of the burst length $b$.  When $b\alt L$, very
different first-passage properties arise compared to those for pure diffusion
in the interval.  Perhaps the most striking is the conditional exit time to
reach $x=L$, corresponding to a state of infection, which has a non-monotonic
dependence on the starting position $x$.  We compute these first-passage
properties from the backward Kolmogorov equations for the exit probabilities
and exit times~\cite{vk,fpp}.  In the concluding section, we briefly discuss
the corresponding first-passage properties for the bursty birth/death model.
This process accounts for the feature that bursts should occur at a rate that
is proportional to the number of live viruses.  It is naturally to model this
situation by defining the rate at which steps occur to be proportional to the
current position of a bursty walk on the interval.

\section{The Model}

In the bursty random walk, unit-length steps to the left occur with
probability $q$, while long steps (bursts) of length $b$ occur with
probability $p=1-q$ (Fig.~\ref{model}).  We choose $p$ and $q$ so that the
average position of the walk does not change at each step; however, most of
our results are derived for general $p$ and $q$.  The motivation for
considering these hopping probabilities is based on the experimental
observation that viral counts in an organism often remain nearly constant for
time periods much longer than the lifetime of individual viruses.  Such a
near constancy could only arise if an organism produces new viruses (by
bursts) and clears viruses at similar overall rates~\cite{perelson}.

With the constraint that the number of virus particles remains fixed, on
average, the respective probabilities of making a single step to the right
and to the left are
\begin{align}
\label{pq}
  p=\frac{1}{b+1}~,\qquad\qquad q=\frac{b}{b+1}~.
\end{align}
The bursty random walk is confined to the finite interval $[0,L]$, where the
coordinate represents the number of live viruses.  The state where the virus
is cleared is represented by the point $x=0$, while the state where a
sufficient number of viruses exists that the organism is ill or dead is
represented by $x=L$.  Our goal is to understand first-passage properties of
this bursty walk that are relevant to the state of health of the organism.
Namely, what is the probability that the organism becomes cured or becomes
ill, corresponding to the walk eventually exiting through the left edge or
the right edge of the interval, respectively?  What is the time needed for
the organism to become cured or ill?

To set the stage for our results, let us recall some well-known first-passage
properties for the isotropic nearest-neighbor random walk on the
interval~\cite{fpp}.  Let $\mathcal{E}_+(x)$ denote the probability that this
random walk eventually exits the interval at $L$ without ever touching $x=0$,
given that the walk started at an arbitrary point $x$, with $0<x<L$.  The
complementary exit probability to the left boundary is
$\mathcal{E}_-(x)=1-\mathcal{E}_+(x)$.  These exit probabilities satisfy the
recursion
\begin{equation}
\label{E}
\mathcal{E}_\pm(x)= \tfrac{1}{2}\,\mathcal{E}_\pm(x-1)+ \tfrac{1}{2}\,\mathcal{E}_\pm(x+1)\,,
\end{equation}
subject to the boundary conditions $\mathcal{E}_+(0)=0,~ \mathcal{E}_+(L)=1$,
or $\mathcal{E}_-(0)=1,~ \mathcal{E}_-(L)=0$.  This recursion expresses the
exit probability starting at $x$ as the probability of first taking a step to
the left or right (the factor 1/2) and then exiting from either $x-1$ or
$x+1$, respectively.  The solution to Eq.~\eqref{E} with these boundary
conditions is:
\begin{equation}
\label{E-classic}
\mathcal{E}_+(x)=\frac{x}{L}~,\qquad \mathcal{E}_-(x) =1-\frac{x}{L}~.
\end{equation}

We also define $t(x)$ as the average time for the walk to leave the interval
at either end when it starts at $x$.  This {\em unconditional\/} exit time
satisfies the recursion
\begin{equation}
\label{t}
t(x)= \tfrac{1}{2}\,t(x-1)+ \tfrac{1}{2}\,t(x+1)+1\,,
\end{equation}
subject to the boundary conditions $t(0)=t(L)=0$.  Equation~\eqref{t}
expresses the average exit time from $x$ as the time for the first step (the
additive factor 1) plus the exit time from the new positions (either $x\pm
1$); the factor $\frac{1}{2}$ accounts for the probability for each of these
two choices.  Similarly, we also define the {\em conditional\/} exit times,
$t_\pm(x)$, as the average times for the walk to leave the interval by the
right or the left boundary, respectively, without ever reaching the opposite
boundary.  These conditional exit times satisfy~\cite{fpp}
\begin{equation}
\label{tpm}
\mathcal{C}_\pm(x)= \tfrac{1}{2}\,\mathcal{C}_\pm(x\!-\!1)
+\tfrac{1}{2}\,\mathcal{C}_\pm(x\!+\!1)+\mathcal{E}_\pm(x)\,,
\end{equation}
with $\mathcal{C}_\pm\equiv \mathcal{E}_\pm t_\pm$, and this equation is
subject to the boundary conditions $\mathcal{C}_\pm(0)=\mathcal{C}_\pm(L)=0$.
For the nearest-neighbor random walk, the exit times are given by
\begin{align}
\label{t-classic}
\begin{split}
  t(x)&=\tfrac{1}{2}\,x(L-x)\,,\\ 
t_+(x)&=\tfrac{1}{3}(L^2-x^2)\,,\\ 
t_-(x)&=\tfrac{1}{3}(2Lx-x^2)\,.
\end{split}
\end{align}

Our goal is to determine the results analogous to Eqs.~\eqref{E-classic} and
\eqref{t-classic} for the bursty random walk.  As we shall see, first-passage
properties depend only on $b/L$ as long as this ratio is nonzero.

\section{Exit Probabilities}
\label{sec:E}

For the bursty random walk with burst length $b$, we may naturally define two
distinct types of exit probabilities to the right boundary:
\begin{itemize}
\item the {\em total\/} exit probability $\mathcal{E}_+(x)$ that the walk
  eventually reaches {\em any} point at the right boundary or beyond, without
  ever touching the left boundary $x=0$;
\item the {\em restricted\/} exit probability $\mathcal{R}_m(x)$ that the
  walk eventually reaches the specific point $L+m$ (with $0\leq m\leq b-1$),
  without ever touching the left boundary or any other point beyond the right
  boundary.  There are $b$ such distinct restricted exit probabilities,
  $\mathcal{R}_m(x)$, with $m=0,1,\ldots,b-1$.
\end{itemize}
While the total exit probability is most relevant physically, because it
corresponds to the probability of illness for a given level of initial
exposure, the restricted exit probabilities display intriguing features that
stem from the bursty character of the walk~\cite{left}.

The total exit probability to the right boundary satisfies the recursion
\begin{equation}
\label{E-rec}
\mathcal{E}_+(x)=q\mathcal{E}_+(x-1)+p\mathcal{E}_+(x+b)\,,
\end{equation}
that represents the extension of Eq.~\eqref{E} to the bursty random walk.
This recursion expresses exit via the right boundary, when starting from $x$,
either by taking the first step to the left (probability $q$), after which
exit from $x-1$ occurs, or by first stepping to the right (probability $p$),
after which exit from $x+b$ may occur.  This recursion must be supplemented
by the boundary conditions $\mathcal{E}_+(x)=1$ for all $x\geq L$ and
$\mathcal{E}_+(0)=0$.  Namely, a walk that starts at $x\geq L$ has already
exited, while a walk that starts at $x=0$ can never exit via the right
boundary.  The equations for the restricted exit probabilities are similar to
\eqref{E-rec}, but are now subject to the boundary conditions
$\mathcal{R}_m(0)=0$ and $\mathcal{R}_m(L+k)=\delta_{k,m}$.

While the exit probabilities can be obtained by enumerating all random walk
trajectories to the exit point and computing the probabilities for all these
paths, the above recursions provide the same results much more
easily~\cite{vk,fpp}.  We will use different methods to solve
Eqs.~\eqref{E-rec} for short and large burst lengths, and therefore study
these cases separately.

\subsection{Burst lengths $b=2, 3, \ldots$}
\label{E-smallb}

For the first non-trivial case of burst length $b=2$, we solve the
constant-coefficient recursion \eqref{E-rec} by attempting a solution of the
form ${E}_+(x)=\lambda^x$.  This leads to the characteristic equation
$\lambda^3-3\lambda+2=0$, with solutions $\lambda=-2$ and $\lambda=1$ (doubly
degenerate).  Henceforth, we use $\lambda$ to denote the first root of the
characteristic polynomial.  The general solution to Eq.~\eqref{E-rec} thus is
$\mathcal{E}_+(x)=a\lambda^x+bx+c$.  Invoking the boundary conditions, the
total exit probability to the right boundary is
\begin{equation}
\mathcal{E}_+(x)=
\frac{x}{L\!+\!1}+\frac{1}{L\!+\!1}\frac{\left[(\lambda^x\!-\!1)-\frac{x}{L+1}(\lambda^{L+1}\!-\!1)\right]}
{\left[(\lambda^L\!-\!1)-\frac{L}{L+1}(\lambda^{L+1}\!-\!1)\right]}~.
\end{equation}

For the restricted exit probabilities to a specific point in the range
$[L,L+b-1]$, the boundary conditions are:
\begin{align*}
&\mathcal{R}_0(0)\!=\!
\mathcal{R}_0(L\!+\!1)\!=\!0,~ \mathcal{R}_0(L)\!=\!1,~\mathrm{exit\ to\ } x=L, \\
&\mathcal{R}_1(0)\!=\! \mathcal{R}_1(L)\!=\!0,~
\mathcal{R}_1(L\!+\!1)\!=\!1\,~~ \mathrm{exit\ to\ } x=L\!+\!1.\nonumber
\end{align*}
Applying these boundary conditions to the general solution $a\lambda^x+bx+c$,
we obtain
\begin{align}
\begin{split}
\mathcal{R}_0(x)&=\frac{(\lambda^x-1)-\frac{x}{L+1}(\lambda^{L+1}-1)}
{(\lambda^L-1)-\frac{L}{L+1}(\lambda^{L+1}-1)} ~, \\
\mathcal{R}_1(x)&=\frac{(\lambda^x-1)-\frac{x}{L}(\lambda^{L}-1)}
{(\lambda^{L+1}-1)-\frac{L+1}{L}(\lambda^{L}-1)}~,
\end{split}
\end{align}
for the restricted exit probabilities to $x=L$ and to $x=L+1$, respectively.
Parenthetically, once we know one of $\mathcal{R}_0(x)$ or
$\mathcal{R}_1(x)$, the other is determined by the martingale property that
the mean position of the bursty walk always remains fixed~\cite{GS01}.  That
is, after all the probability has reached an absorbing boundary, the two
restricted exit probabilities are related by $ 0\times [1-
\mathcal{R}_0(x)-\mathcal{R}_1(x)]+L\times \mathcal{R}_0(x)+ (L+1)\times
\mathcal{R}_1(x)=x$.  The restricted exit probabilities initially grow nearly
linearly in $x/L$ (Fig.~\ref{b23}), but then oscillate violently as $x\to L$.
The total exit probability $\mathcal{E}_+(x)$ is a nearly linear function of
$x$ for small $x$ but its slope develops oscillations as $x\to L$.

This same calculational method can be extended to longer bursts.  By assuming
an exponential solution of the form $\mathcal{E}_+(x)=\lambda^x$ in
Eq.~\eqref{E-rec}, the characteristic polynomial generically is
$(\lambda-1)^2A(\lambda)$, where $A(\lambda)$ is a polynomial of order $b-1$.
Explicit closed-form solutions can therefore be obtained for $b\leq 5$, but
numerically exact results can be obtained for any burst length; details for
the case $b=3$ are given in appendix~\ref{app:shortb}.  Typical results are
shown in Fig.~\ref{b23} for burst lengths $b=2$, 10, and also
$b=\frac{L}{3}$, $\frac{L}{2}$, and $L$.  The total exit probability is very
close to linear function with slope less than one for $x<L-b$, but deviates
from linearity within one burst length from $x=L$.  The restricted exit
probabilities are also nearly linear functions for $x<L-b$, but oscillate
violently in the boundary region.

\subsection{Long Bursts}

When the burst length is of the order of the interval length, we can simplify
the determination of the exit probabilities by considering separate
recursions in each of the disjoint subintervals $[L-b,L]$, $[L-2b,L-b-1]$,
$[L-3b,L-2b-1]$, etc., instead of directly solving for the roots of a
characteristic polynomial of order $b-1$.  As we shall see, this partitioning
significantly reduces the order of the recursions for the exit probabilities.


\subsubsection{Total exit probabilities}

\begin{figure}[ht]
  \includegraphics*[width=0.35\textwidth]{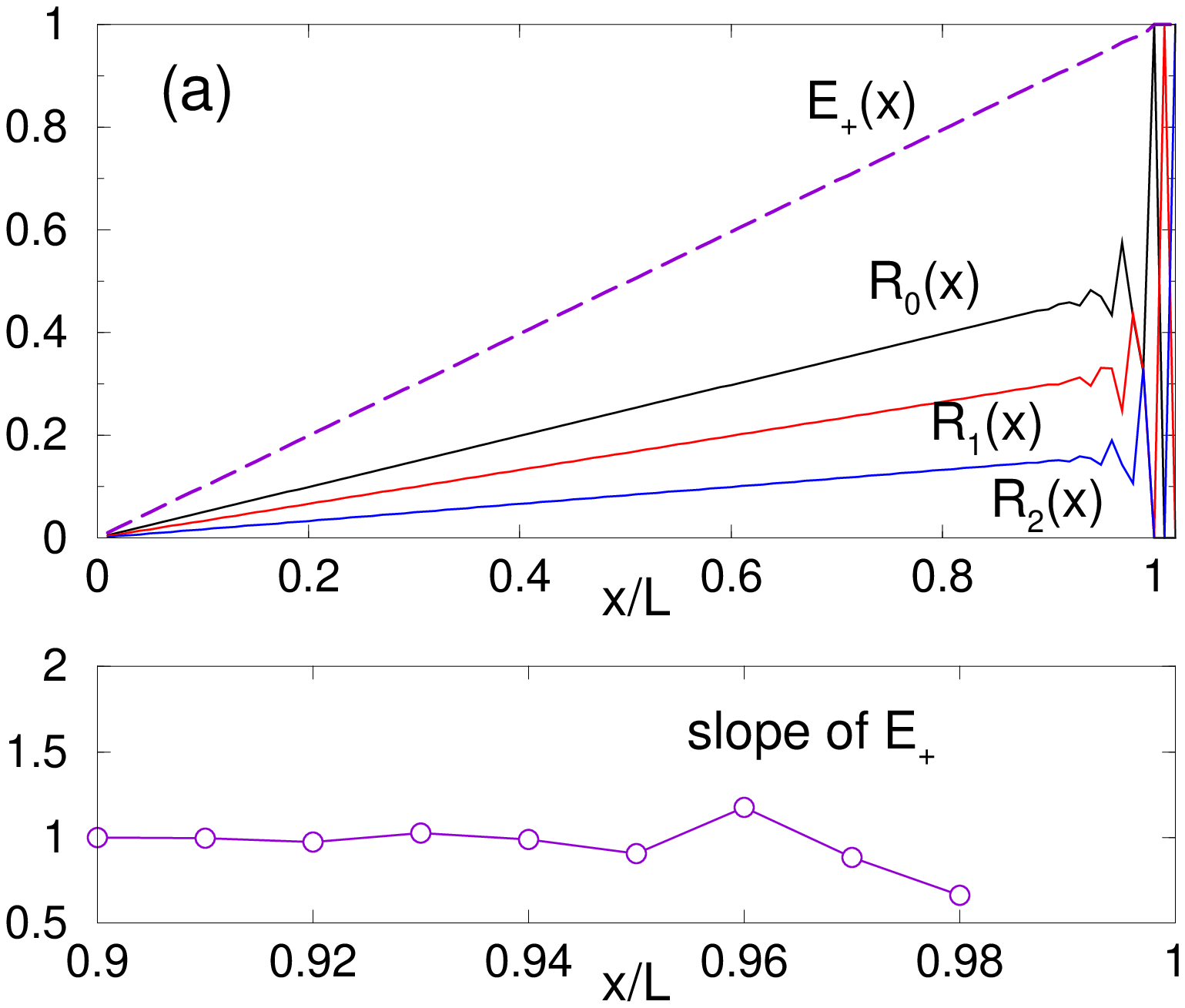}
\includegraphics*[width=0.35\textwidth]{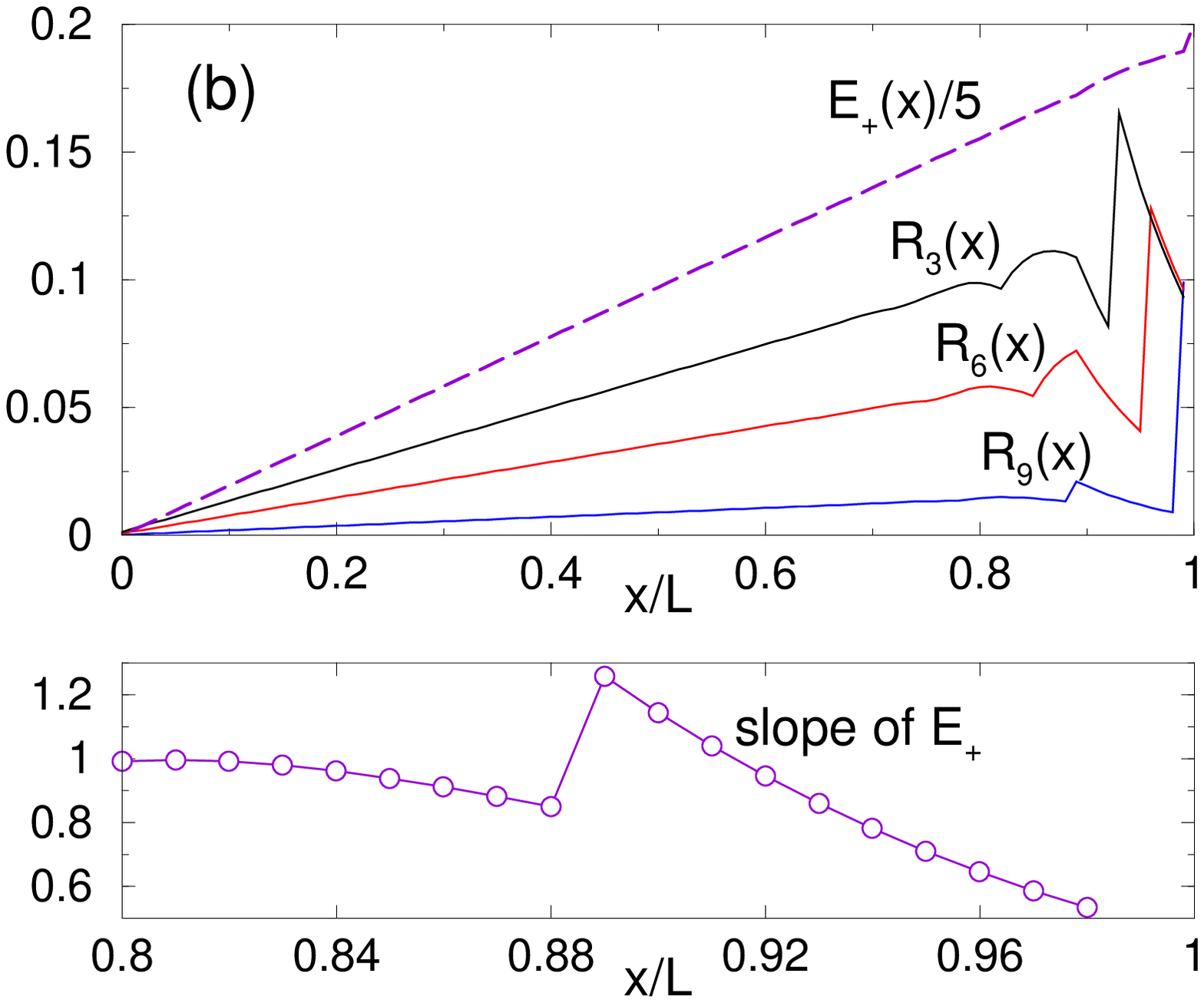}
\includegraphics*[width=0.35\textwidth]{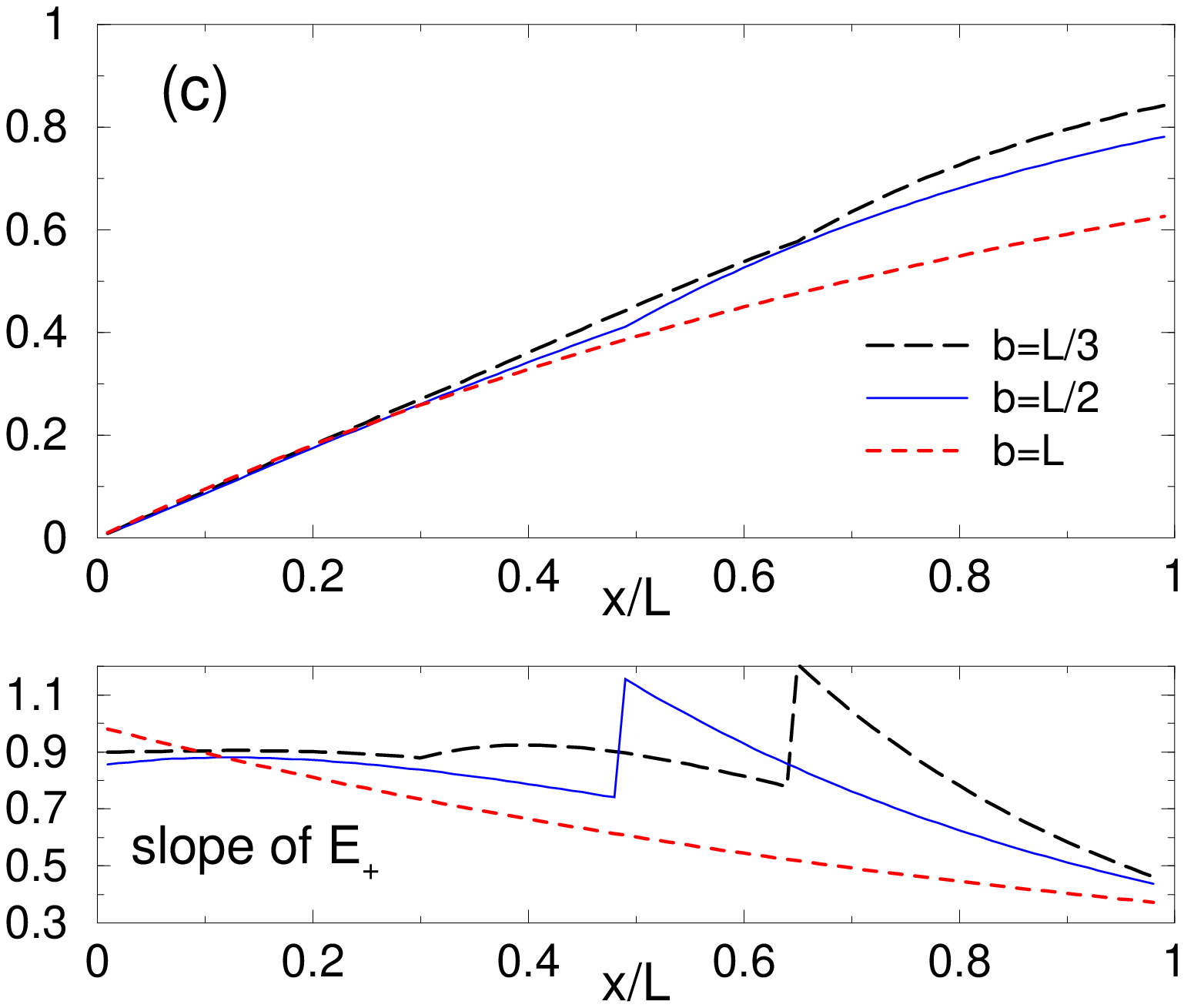}
\caption{Exit probabilities for the bursty random walk for: (a) burst length
  $b=3$, (b) $b=10$, and (c) $b=L$, $L/2$, and $L/3$.  Simulations are on a
  system of length $L=100$.}
  \label{b23}
\end{figure}

In the extreme situation where the burst length $b\geq L$, a single burst
results in exit at or beyond the right end of the interval.  Thus the total
exit probability satisfies the recursion
$\mathcal{E}_+(x)=q\mathcal{E}_+(x-1)+p$; that is, either the walk steps to
the left and then exits from $x-1$, or the walk steps to the right and exits
immediately.  The solution to this recursion is a constant plus an
exponential function.  The boundary condition $\mathcal{E}_+(0)=0$
immediately gives
\begin{equation}
\label{EL}
  \mathcal{E}_+(x)=1-q^x.
\end{equation}
Because of the overwhelming probability of stepping to the left, the exit
probability to the right boundary is not close to one for $x\to L$ from
below.  As an example, for $b=L$, we have $\mathcal{E}_+(L-1)\to
1-e^{-1}\approx 0.6321$ (Fig.~\ref{b23}(c)).

\begin{figure}[ht]
  \centerline{\includegraphics*[width=.4\textwidth]{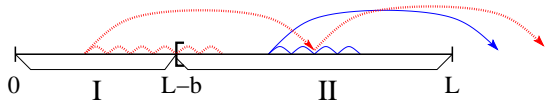}}
  \caption{Partitioning of the interval into $[0,L-b-1]$ (region I)
    and $[L-b,L]$ (region II).  To exit from region I requires at least two
    bursts.}
  \label{partition}
\end{figure}

For the case $L/2\leq b<L$, we partition $[0,L]$ into the subintervals
$[0,L-b-1]$ (defined as region I) and $[L-b,L]$ (region II), as indicated in
Fig.~\ref{partition}.  Making a slight abuse of notation, we define
$\mathcal{E}^{\rm I}(x)$ and $\mathcal{E}^{\rm II}(x)$ as the total exit
probabilities to $x\geq L$, when starting at a point $x$ that is in either
region I or region II, respectively.  These exit probabilities satisfy
\begin{align}
\begin{split}
\label{2-int}
\mathcal{E}^{\rm I}(x)&=q\mathcal{E}^{\rm I}(x-1)+p\mathcal{E}^{\rm II}(x+b)\,,\\
\mathcal{E}^{\rm II}(x)&=q\mathcal{E}^{\rm II}(x-1)+p\,.
\end{split}
\end{align}
These recursions are identical in form to Eqs.~\eqref{E-rec}, but with the
subinterval explicitly identified.  Thus, from example, exit to $x\geq L$,
when starting from a point $x$ in region I, can occur by taking a step to the
left with probability $q$ and then exiting from $x-1$ (necessarily in region
I), or by taking a step to the right with probability $p$ and then exiting
from $x+b$ (necessarily in region II).  Equations~\eqref{2-int} are subject
to the boundary condition $\mathcal{E}^{\rm I}(0)=0$ as well as the joining
condition $\mathcal{E}^{\rm II}(L-b)=q\mathcal{E}^{\rm I}(L-b-1)+p$.

By this partitioning, the exit probabilities in each subinterval are
functionally distinct and can be solved separately.  In the second of
Eqs.~\eqref{2-int}, a particular solution is $\mathcal{E}^{\rm II}_{\rm
  par}=1$.  Thus the general solution has the form $\mathcal{E}^{\rm
  II}(x)=1+Aq^x$.  Substituting this expression in the first of
Eqs.~\eqref{2-int}, now gives the closed recursion $\mathcal{E}^{\rm
  I}(x)=q\mathcal{E}^{\rm I}(x-1)+p+Apq^x$.  With the inhomogeneous term
$p+Apq^x$, the general solution is $\mathcal{E}^{\rm I}(x)=A+(Bx+C)q^x$.
Using the boundary condition $\mathcal{E}^{\rm I}(0)=0$, and substituting
this form for $\mathcal{E}^{\rm I}(x)$ into the first of Eqs.~\eqref{2-int},
we find $B=Apq^b$.  Finally, we invoke the joining condition and obtain
\begin{equation}
\label{E2-result}
  \mathcal{E}^{\rm I}(x)= 1\!-\! q^x-\frac{x\,p\, q^{x+b}}{1\!-\!ypq^b}~,\quad
  \mathcal{E}^{\rm II}(x)= 1\!-\!\frac{ q^x}{1\!-\!ypq^b}~,
\end{equation}
where $y\equiv L-b-1$.  Notice again that because of the large probability of
hopping to the left, $\mathcal{E}_{\rm II}(x)$ is discontinuous as $x\to L$.

For $L/3\leq b<L/2$, we partition $[0,L]$ into the three subintervals
$[0,L-2b-1]$, $[L-2b,L-b-1]$, and $[L-b,L]$, (regions I, II, and III
respectively) and solve the generalization of Eqs.~\eqref{2-int} to three
intervals, supplemented by two joining conditions at $x=L-b$ and at $x=L-2b$
(appendix~\ref{app:longb}).  As shown in Fig.~\ref{b23}, the total exit
probability has two (barely visible) singularities and deviates considerably
from linearity within one burst length from the right boundary.  Generally,
for a partitioning into $k$ intervals, the slope of the total exit
probability is discontinuous at the boundary between intervals $k$ and $k-1$,
the second derivative is discontinuous at the boundary between intervals
$k-1$ and $k-2$, the third derivative is discontinuous at the boundary
between intervals $k-2$ and $k-3$, etc.  A similar intricate pattern of a
sequence of progressively weaker singularities arises in various
fragmentation models~\cite{DF87,FIK95}.

\subsubsection{Restricted exit probabilities}

The restricted exit probabilities to a specific point undergo a more dramatic
sequence of discontinuities between successive subintervals.  We again start
with the case where $b$ lies in the range $[L/2,L]$ so that there are two
subintervals to consider: $[0,L-b-1]$ and $[L-b,L]$.  For concreteness we
determine the exit probability to the specific site $x=L$; similar behavior
arises for other exit points in $[L,L+b-1]$.  Now the recursion relations for
the restricted exit probabilities are
\begin{align}
\begin{split}
\label{int-2r}
\mathcal{R}^{\rm I}(x)&=q\mathcal{R}^I(x-1)+p\mathcal{R}^{\rm II}(x+b)\,,\\
\mathcal{R}^{\rm II}(x)&=q\mathcal{R}^{\rm II}(x-1)\,.
\end{split}
\end{align}
Since we seek only the exit probability to $x=L$, we simplify notation by
omitting the subscript that specifies the exit location; thus
$\mathcal{R}_0\to\mathcal{R}$.  The first equation states that to reach $x=L$
from subinterval I, the walk can either step left (probability $q$) and exits
from $x-1$, or the walk steps to the right (probability $p$) and exits from
$x+b$.  The second equation states that to reach $x=L$ from within
subinterval II, the only possibility is to step to the left; a burst would
lead to exit at a point $x>L$, which does do not contribute to the exit
probability to $x=L$.  The recursions \eqref{int-2r} must be supplemented by
the boundary condition $\mathcal{R}^{\rm I}=0$ and the joining condition
$\mathcal{R}^{\rm II}(L-b)=q\mathcal{R}^{\rm I}(L-b-1)+p$.  Notice that exit
to $L$ can occur {\em only} if the walk is at the point $x=L-b$.

\begin{figure}[ht!]
  \centerline{\includegraphics*[width=.23\textwidth]{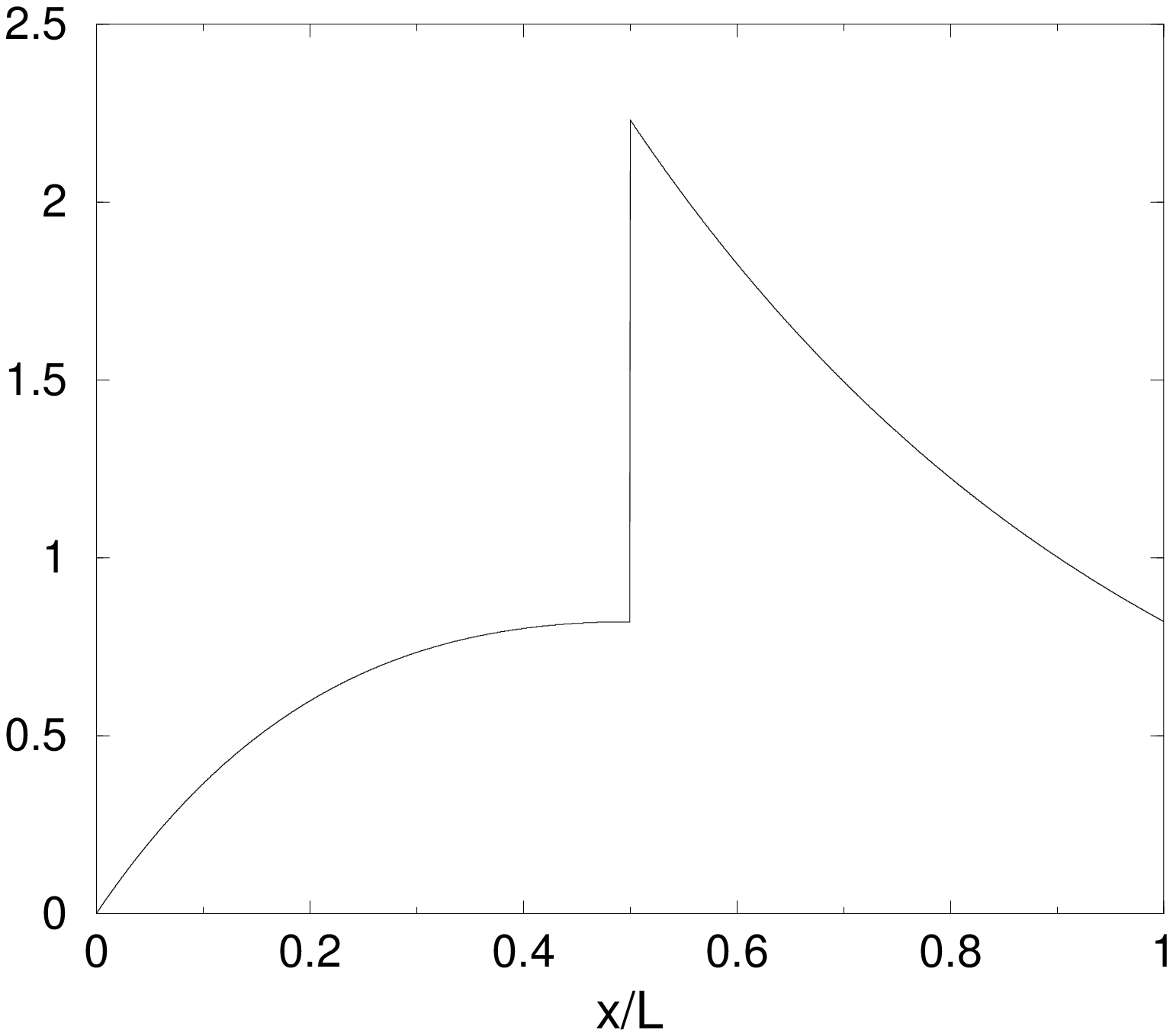}\quad\includegraphics*[width=.23\textwidth]{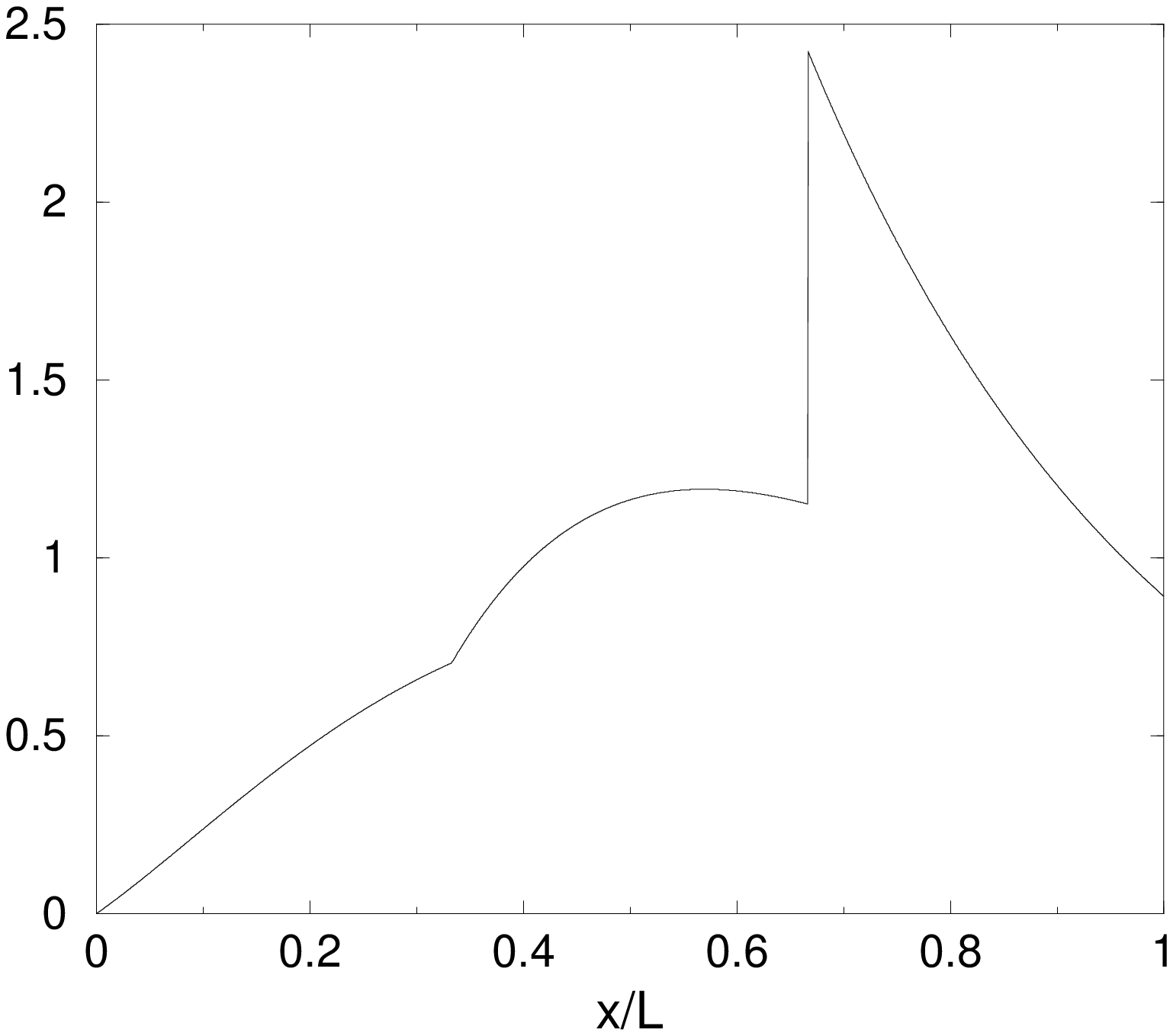}}
  \caption{Restricted exit probabilities to $x=L$ for the cases of $b=L/2$
    and $b=L/3$.  The arbitrary vertical scale has been fixed by setting the
    integral of $\mathcal{R}(x)$ over the interval equal to 1.}
  \label{intervals}
\end{figure}

Employing the same method as that used to obtain Eqs.~\eqref{E2-result}, we
now obtain obtain
\begin{align}
  \mathcal{R}^{\rm I}(x)=\frac{xp^2 q^{x+2 b-L}}{1-yp q^b}~,\qquad
  \mathcal{R}^{\rm II}(x)=\frac{p q^{x+b-L}}{1-yp q^b}~.
\end{align}
This solution method can be extended to more subintervals; the results for
$b=L/2$ (two intervals) and $b=L/3$ (three intervals) are shown in
Fig.~\ref{intervals}.  For a partitioning into $k$ intervals, the exit
probability is discontinuous at the boundary between intervals $k$ and $k-1$,
the first derivative is discontinuous at the boundary between intervals $k-1$
and $k-2$, etc.; the pattern is similar to that for the total exit
probability, but the discontinuities are more prominent here since they begin
with the function itself rather than with the first derivative.

\section{First-Passage Times}
\label{sec:t}

By adapting Eq.~\eqref{t} to the bursty random walk, the unconditional mean
first-passage time satisfies
\begin{equation}
\label{T-rec}
t(x)=qt(x\!-\!1)+pt(x\!+\!b)+1\,,
\end{equation}
subject to the boundary conditions $t(0)=0$ and also $t(L\!+\!m)=0$ for
$m=0,1,\ldots b-1$.  Similarly, the quantities $\mathcal{C}_\pm(x)$, which
are related to the conditional exit times, satisfy the recursion (see
Eq.~\eqref{tpm})

Similarly, the conditional mean first-passage times
$\mathcal{C}_\pm(x)$ satisfy the recursion (see Eq.~\eqref{tpm})
\begin{equation}
\label{f+}
\mathcal{C}_\pm(x)=q\mathcal{C}_\pm(x\!-\!1)+p\mathcal{C}_\pm(x\!+\!b)+\mathcal{E}_\pm(x)\,,
\end{equation}
subject to the same boundary conditions as for $t(x)$ itself.  Again, we
treat the exit times separately for short and for long bursts.

\subsection{Short bursts}

For the first non-trivial case of $b=2$, let us focus on the unbiased case of
$p=\frac{1}{3}$ and $q=\frac{2}{3}$ for simplicity.  We solve the recursion
\eqref{T-rec} with these values of $p$ and $q$ by noting that the
inhomogeneous term can be eliminated by writing $t(x)=T(x)-x^2/2$.
Substituting this ansatz into Eq.~\eqref{T-rec}, we find that $T(x)$ obeys
this same equation, but without the inhomogeneous term.  From our analysis of
the exit probability in Sec.~\ref{E-smallb}, the general solution is
$T(x)=a\lambda^x+bx+c$, with $\lambda=-2$, subject to the boundary conditions
$T(0)=0$, $T(L)=L^2/2$, $T(L+1)=(L+1)^2/2$ that correspond to
$t(0)=t(L)=t(L+1)=0$.  We thereby obtain, for the unconditional mean
first-passage time,
\begin{equation}
t(x)=\frac{1}{2}x(L\!-\!x) +\frac{1}{2}\frac{(L\!+\!1)\big[(\lambda^x\!-\!1)-\frac{x}{L}(\lambda^L\!-\!1)\big]}{(\lambda^{L+1}-1)-\frac{L+1}{L}(\lambda^L\!-\!1)}~,
\end{equation}
with $\lambda=-2$.  The second term represents a tiny correction to the
leading diffusive behavior of $\frac{1}{2}x(L-x)$.  

\subsection{Long Bursts}

\begin{figure}[ht!]
\includegraphics*[width=0.3125\textwidth]{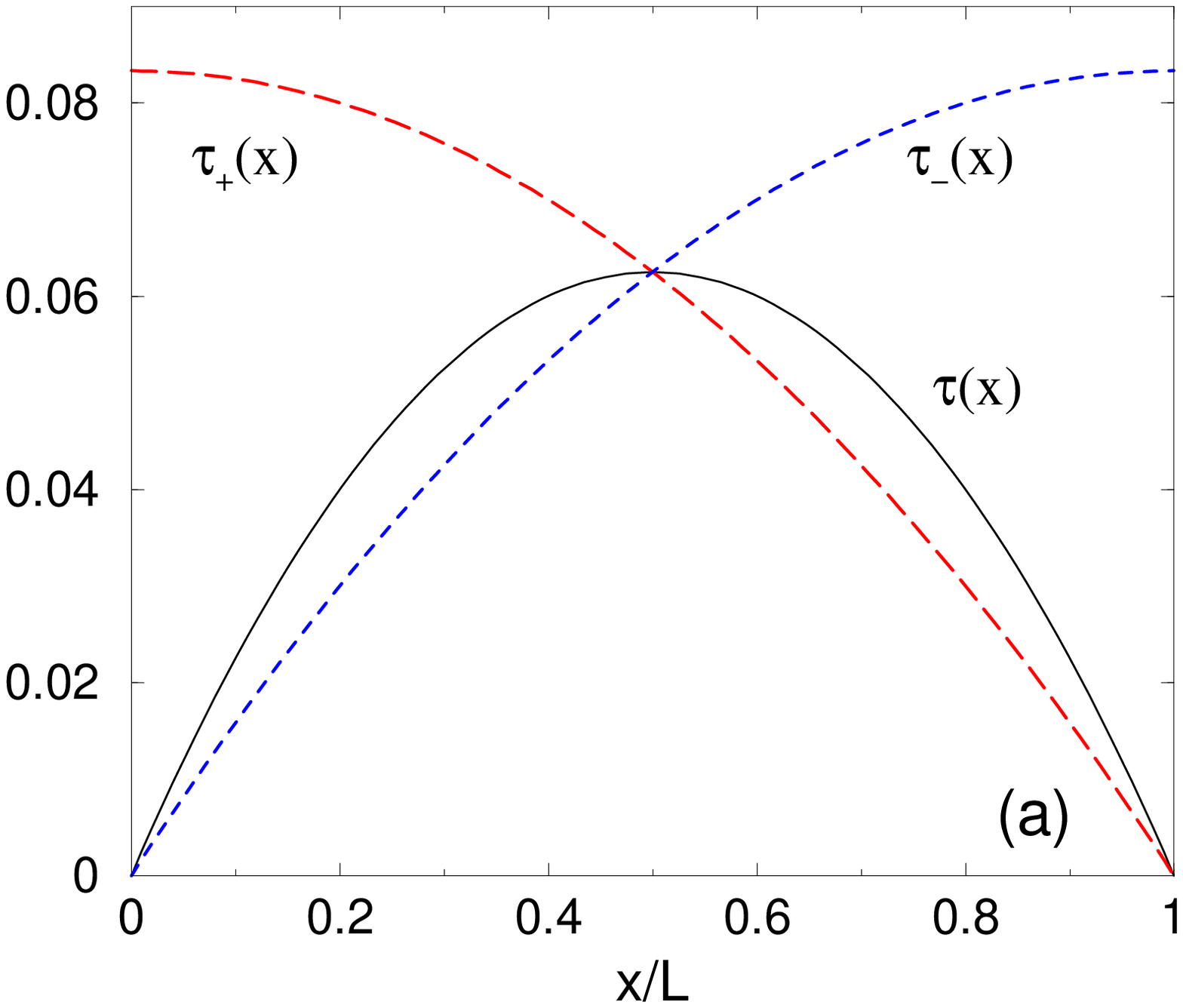}
\includegraphics*[width=0.3125\textwidth]{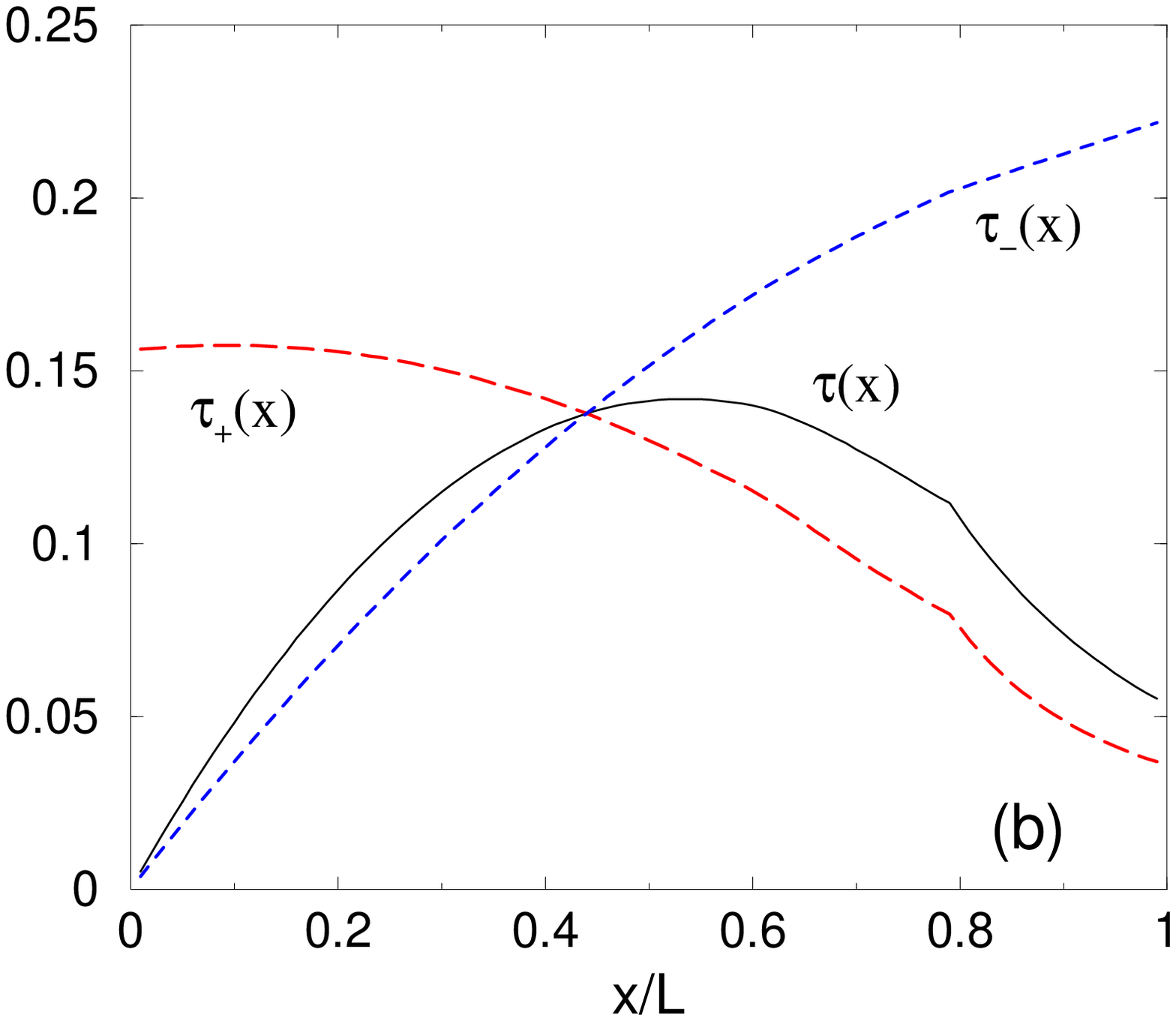}
\includegraphics*[width=0.3125\textwidth]{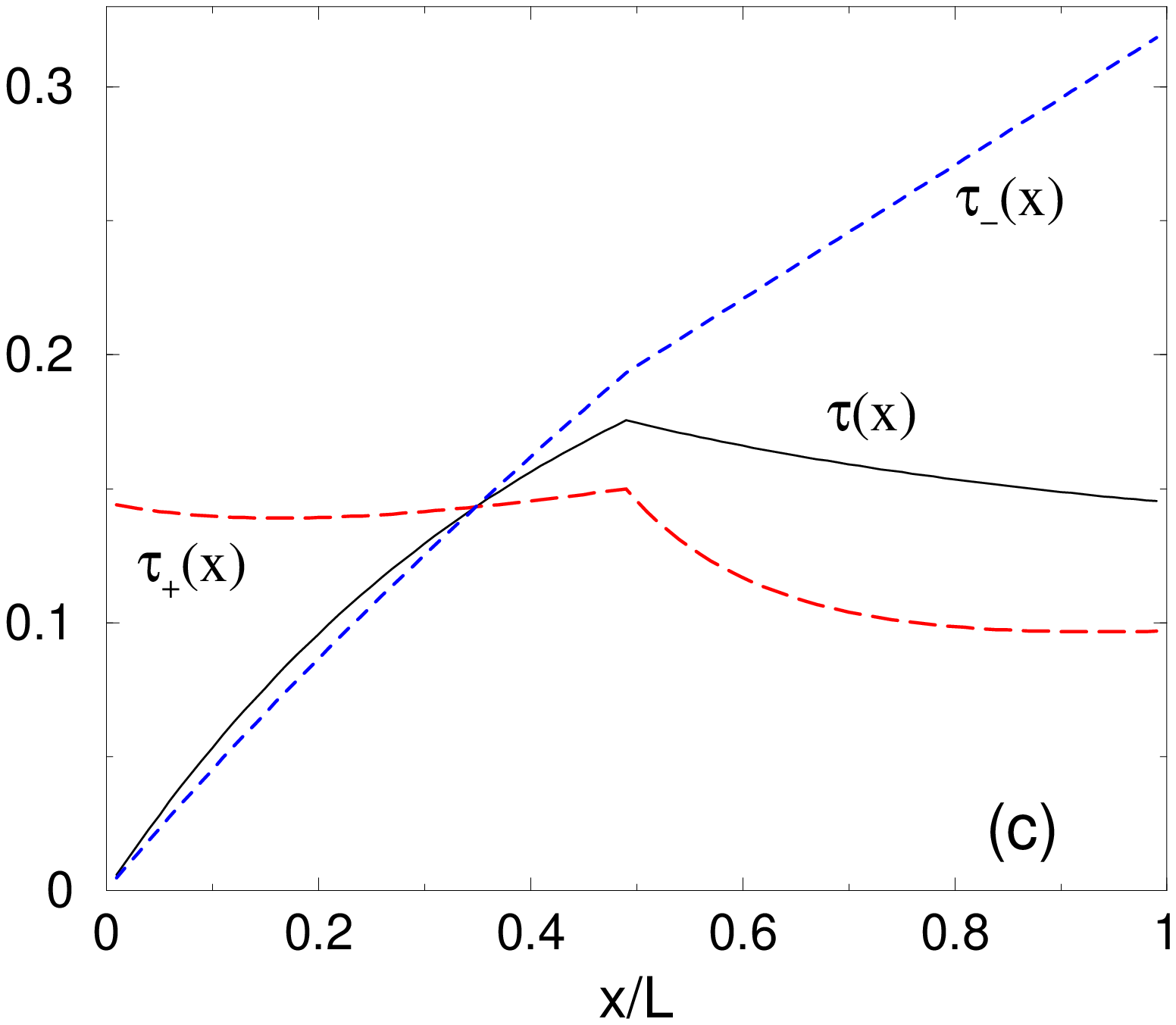}
\includegraphics*[width=0.3125\textwidth]{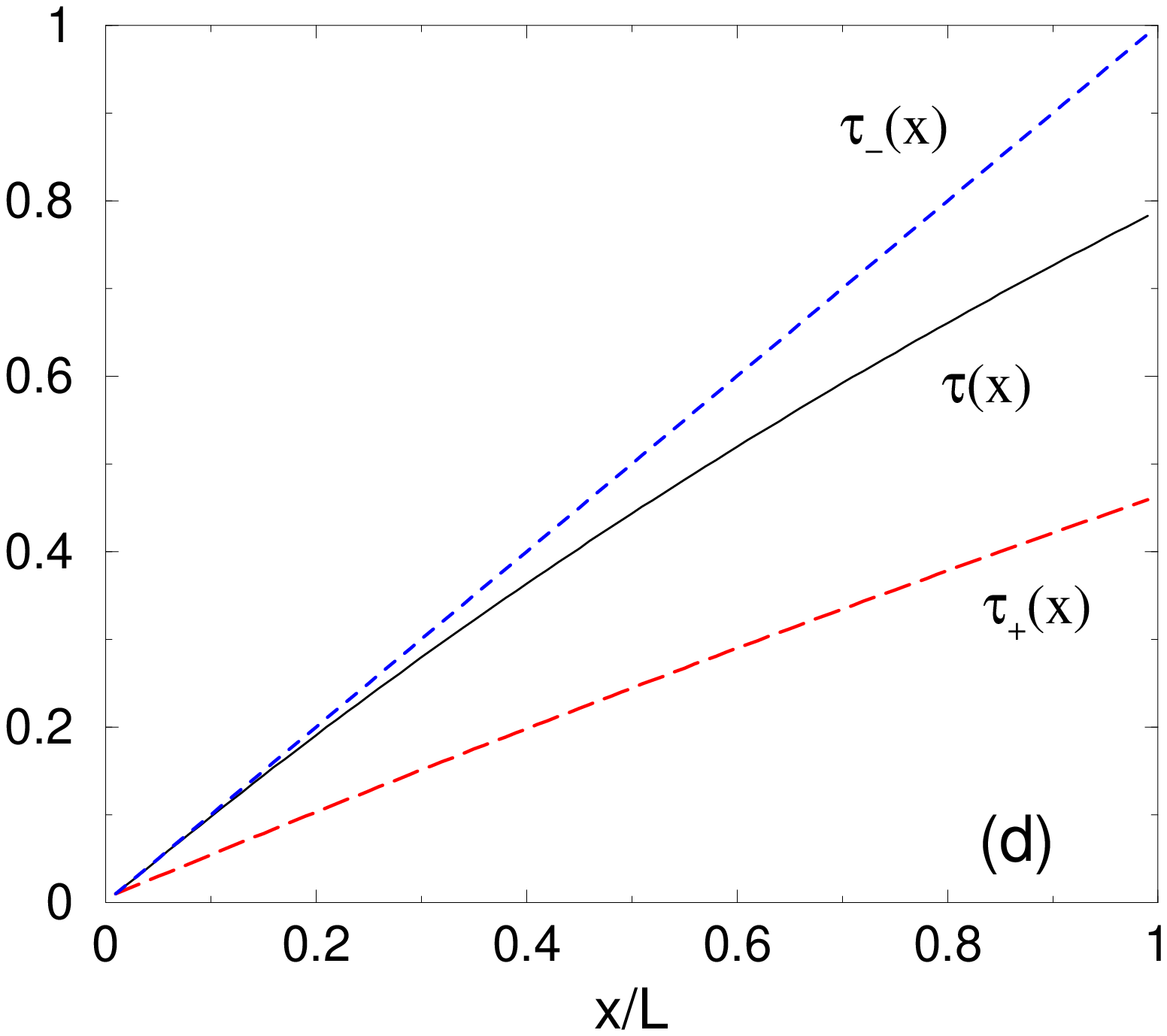}
\caption{ Normalized mean first-passage times $\tau \equiv t/(L^2\!/\!D)$,
  with $D=b/2$, for: (a) the nearest-neighbor random walk, and the bursty
  random walk with: (b) burst length $b=L/5$, (c) $b=L/2$, and (d) $b=2L$.
  Shown are the unconditional first-passage time $\tau(x)$ and the
  conditional times, $\tau_\pm(x)$, to the right and left boundary,
  respectively. }
  \label{mfpt}
\end{figure}

In the extreme case of burst length $b\geq L$, the walk exits after any
single burst, and the unconditional first-passage time satisfies
$t(x)=qt(x-1)+1$, subject to the boundary condition $t(0)=0$.  The solution
is
\begin{align}
t(x)=\frac{1-q^x}{1-q~}~.
\end{align}
Similarly, the conditional mean first-passage time to the right boundary,
$t_+=\mathcal{C}_+/\mathcal{E}_+$, is determined from the recursion
\begin{equation}
\mathcal{C}_+(x)=q\mathcal{C}_+(x-1)+\mathcal{E}_+(x)=q\mathcal{C}_+(x-1)+1-q^x\,,
\end{equation}
subject to the boundary condition $\mathcal{C}_+(0)=0$.  The solution now is
\begin{align}
  t_+(x)=\frac{1}{1-q}-\frac{x\,q^x}{1-q^x}~.
\end{align}
The conditional exit time $t_-$ may be obtained from the conservation
statement $t(x)=\mathcal{E}_-(x)t_-(x)+\mathcal{E}_+(x)t_+(x)$ and gives
$t_-(x)=x$.  An apparently paradoxical feature is that the exit time $t_+$
increases when the starting point is closer to $x=L$ (Fig.~\ref{mfpt}(d)).
This behavior arises because steps to the left occur with overwhelming
probability.  Thus a walk that starts near $x=L$ will almost surely hop a
considerable distance to the left before a burst occurs.  However a walk that
starts near $x=0$ can only hop a short distance to the left before a burst
must occur to ensure exit at the right boundary.

For $L/2\leq b<L$, we again partition the interval into the subintervals
$[0,L-b-1]$ (region I) and $[L-b,L]$ (region II) and denote the mean
first-passage times within each as $t^{\rm I}(x)$ and $t^{\rm II}(x)$
respectively.  The unconditional mean first-passage time satisfies
\begin{align}
\begin{split}
t^{\rm I}(x)&=qt^{\rm I}(x-1)+pt^{\rm II}(x+b)+1\,,\\
t^{\rm II}(x)&=qt^{\rm II}(x-1)+1\,,
\end{split}
\end{align}
subject to the boundary condition $t^{\rm I}(x)=0$ and the joining condition
$t^{\rm II}(L-b)=qt^{\rm I}(L-b-1)+1$.  Solving first for $t^{\rm II}$ and
then using this solution in the equation for $t^{\rm I}$, we obtain
\begin{align}
\begin{split}
t^{\rm I}(x)&=\frac{x\left(q^{-y}-2\right) q^{b+x}}{1-ypq^b}-\frac{2 \left(q^x-1\right)}{p}~,
 \\
t^{\rm II}(x)&=\frac{\left(1-ypq^b+q^{x-y}-2q^x\right)}{p\left(1-ypq^b\right)}~,
\end{split}
\end{align}
with $y=L-b-1$.

For the conditional first-passage time to the right boundary, the quantity
$\mathcal{C}_+(x)=\mathcal{E}_+(x)t_+(x)$ satisfies
\begin{align}
\begin{split}
\mathcal{C}^{\rm I}(x)&=q\mathcal{C}^{\rm I}(x-1)+p\mathcal{C}^{\rm II}(x+b)+\mathcal{E}^{\rm I}(x)\,,\\
\mathcal{C}^{\rm II}(x)&=q\mathcal{C}^{\rm I}(x-1)+\mathcal{E}^{\rm II}(x)\,,
\end{split}
\end{align}
subject to the boundary condition $\mathcal{C}^{\rm I}(x)=0$ and the joining
condition $\mathcal{C}^{\rm II}(L-b)=q\mathcal{C}^{\rm
  I}(L-b-1)+\mathcal{E}^{\rm II}(L-b)$.  Again, we have made the notational
abuse of dropping the subscript $\pm$ and focusing only on the exit time to
the right boundary.  Solving these equations for $\mathcal{C}_+$ and dividing
by $\mathcal{E}_+(x)$ yields the conditional first-passage time to the right
boundary (Fig.~\ref{mfpt}).  This same calculation can be straightforwardly
(but tediously) extended to smaller values of $b$, corresponding to more
subintervals.

\begin{figure}[ht!]
  \centerline{\includegraphics*[width=0.35\textwidth]{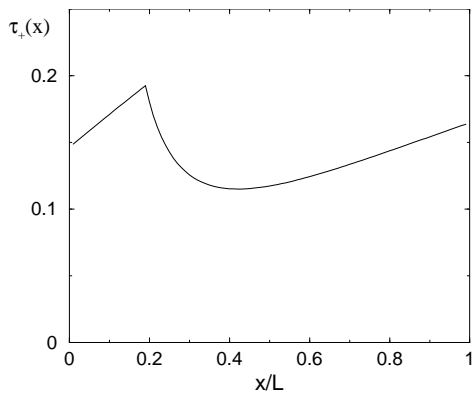}}
  \caption{The normalized conditional exit time to the right boundary,
    $\tau_+ \equiv t_+/(L^2\!/\!D)$ for the case $b/L=0.8$.}
  \label{t80}
\end{figure}

A peculiar feature of the conditional first-passage time $t_+(x)$ is its
non-monotonic dependence on $x$ as the burst length becomes of the order of
the system length (Fig.~\ref{t80}).  This non-monotonicity has a simple
origin.  For $x/L\alt 1$, a typical walk will move a considerable distance to
the left before exit occurs.  Thus, in some sense, points near the right
boundary are ``further'' from the exit than points in the interior of the
interval.  Similarly, a particle that starts near $x=0$ must quickly hop to
the right to avoid exiting at the left boundary.  Thus again, the exit time
to the right is an increasing function of $x$ in this range.  Finally, for a
particle that starts in a narrow range in which $x$ is slightly larger than
$L-b$, the exit time decreases as $x$ increases.  The source of this
decreasing dependence on $x$ in this range is that a particle with $x\agt
L-b$ is increasingly likely to reach a point that is less than $L-b$ as $x$
decreases toward $L-b$.  Once the point $x=L-b$ is crossed, two bursts are
required for exit to the right and typically there will be many steps to the
left between these two bursts.  Thus the exit time increases rapidly as the
starting point approaches $L-b$ from above.

\section{Discussion}

We investigated the first-passage properties of the bursty random walk on a
finite interval, where short steps to the left occur with a high probability,
while long steps to the right --- ``bursts'' --- occur with a small
probability.  The disparity in these hopping probabilities is needed to
ensure that there is no net displacement of a random walker, a feature that
maximizes the time for the walker to survive within the interval.  This model
was motivated by the problem of the early stages of virus spread after
initial exposure~\cite{PKP10}.

When the burst length is short, there are only small corrections to the
well-known first-passage properties of the nearest neighbor random walk.
Conversely, when the burst length is of the order of the interval length,
discreteness effects play an important role.  For such burst lengths, we
solved for first-passage properties by partitioning the full interval into
disjoint subintervals of length $b$, solving each one separately, and then
patching together these subinterval solutions by invoking appropriate joining
conditions.  Strikingly, the mean first-passage time to the right boundary,
corresponding to the time for a host organism to become ill, has a
non-monotonic dependence on the initial location for $b/L\alt1$
(Fig.~\ref{t80}).  Another basic feature of the first-passage properties for
large $b$ is that they are functions of $b/L$ rather then depending on $b$
and $L$ separately.

In spite of the strange behavior of the mean first-passage time, the
distribution of first-passage times is generically characterized by an
exponential decay, but with superimposed oscillations due to burstiness.
Consequently, higher moments of the first-passage times can be simply
characterized by powers of the first moment.

\begin{figure}[h]
 \centerline{\includegraphics*[width=0.35\textwidth]{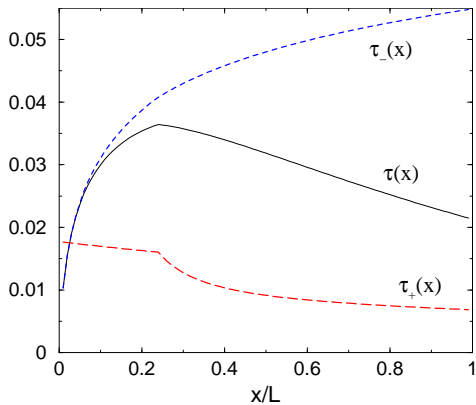}}
 \caption{The normalized unconditional and conditional exit times,
   $\tau\equiv t/L$, for the bursty birth/death process with $b/L=0.75$.}
 \label{bd75}
\end{figure}

If one takes seriously the equivalence that the position of the random walker
as equivalent to the number of active viruses, then the frequency of bursts
as well as the frequency of virus death events should also be proportional to
the current position of the walk.  Thus it would be more realistic to
consider the bursty birth/death process, where the rate at which the random
walker hops is proportional to its current location.  If a step does occur,
then  a unit-length step to the left occurs with probability $q$ and
a step of length $b$ to the right occurs with probability $p\ll q$.

Because the exit probabilities are independent of the rate at which steps
occur, all our results about exit probabilities continue to hold for the
bursty birth/death process.  However, exit times for bursty birth/death are
quite different from those of the bursty random walk.  For example, the 
unconditional exit time for bursty birth/death satisfies the recursion 
\begin{equation}
\label{t-bd}
t(x)=qt(x-1)+qt(x+b)+\delta t(x)\,,
\end{equation}
where $\delta t(x)$, the microscopic time step at position $x$, is
proportional to ${1}/{x}$.  For the classic birth/death process (burst
length $b=1$) and in the continuum limit, Eq.~\eqref{t-bd} becomes
$t''(x)=-{2}/{x}$ with solution $t(x)=2x\ln(L/x)$.  Over most of the
interval range, this exit time scales linearly with $L$, compared to
$t(x)\sim L^2$ for the exit time of the nearest-neighbor random walk.  For
bursty birth/death, representative results for exit times are given in
Fig.~\ref{bd75}.  While no longer non-monotonic in $x$, the conditional exit
time $t_+(x)$ has a near plateau when the initial position $x<L-b$ and then
decreases in $x$.  Thus once an infection has progressed to a certain
threshold, illness quickly ensues.

\smallskip

We thank Paul Krapivsky, John Pearson, and Alan Perelson for useful advice.
We also gratefully acknowledge financial support from NSF grant DMR0906504.

\begin{widetext}
\appendix

\section{Exit probabilities for burst length $b=3$}
\label{app:shortb}

For burst length $b=3$, the recursion relation for the total exit probability
to the right boundary is
\begin{equation}
\label{app-rec}
\mathcal{E}_+(x)=\tfrac{3}{4}\mathcal{E}_+(x-1)+\tfrac{1}{4}\mathcal{E}_+(x+3)\,.
\end{equation}
Assuming the exponential form $\mathcal{E}_+=\lambda^x$ and substituting into
\eqref{app-rec}, the characteristic equation is $\lambda ^4-4\lambda+3=0$, with
solutions $\lambda=1$ (doubly degenerate) and $\lambda_{\pm}=-1\pm
i\sqrt{2}\equiv\sqrt{3}\, e^{i\phi}$, with $\phi=\tan^{-1}(-\sqrt{2})$.  The
general solution is $\mathcal{E}_+(x)=a\lambda_+^x+b\lambda_-^x+cx+d$.  Now
we impose the boundary conditions $\mathcal{E}_+(0)=0$ and
$\mathcal{E}_+(L)=\mathcal{E}_+(L+1)=\mathcal{E}_+(L+2)=1$ one by one.  The
boundary condition $\mathcal{E}_+(0)=0$ gives
\begin{equation}
\label{app-Einit}
\mathcal{E}_+(x)=a(\lambda_+^x-1)+b(\lambda_-^x-1)+cx\,.
\end{equation}
The boundary condition $\mathcal{E}_+(L)=1$ gives
\begin{eqnarray}
\mathcal{E}_+(x)&=&a\left[(\lambda_+^x-1)-(\lambda_+^L-1)\frac{x}{L}\right]
+b\left[(\lambda_-^x-1)-(\lambda_-^L-1)\frac{x}{L}\right] +\frac{x}{L}~,\nonumber\\
&\equiv&a\alpha(x)+b\alpha^*(x)+\frac{x}{L}~,
\end{eqnarray}
with $\alpha(L)=0$.  Next we impose $\mathcal{E}_+(L+1)=1$ to give
\begin{eqnarray}
\mathcal{E}_+(x) &=&a\big[\alpha(x)\alpha^*(L+1)-\alpha^*(x)\alpha(L+1)\big]
+\frac{x}{L}+\left(1-\frac{x}{L}\right)\frac{\alpha^*(x)}{\alpha^*(L+1)}~,\\
&\equiv& a W(x,L+1)+\frac{x}{L}+\left(1-\frac{x}{L}\right)\frac{\alpha^*(x)}{\alpha^*(L+1)}~,
\end{eqnarray}
where the Wronskian $W(L+1,L+1)=0$.  Finally, imposing $\mathcal{E}_+(L+2)=1$
gives
\begin{equation}
\label{app-E}
\mathcal{E}_+(x)= \frac{W(x,L+1)}{W(L+2,L+1)}\left[1-\frac{L+2}{L}-\frac{1}{L}\frac{\alpha^*(L+2)}{\alpha^*(L+1)}\right]+
\left[\frac{x}{L}-\frac{1}{L}\frac{\alpha^*(x)}{\alpha^*(L+1)}\right]~.
\end{equation}
By inspection, it is clear that Eq.~\eqref{app-E} satisfies all the boundary
conditions; this solution is also real.

A similar calculation can be performed for all the restricted exit
probabilities.  For example, for the restricted exit probability
$\mathcal{R}_0(x)$ to $L$, we start with Eq.~\eqref{app-Einit} and next
impose $\mathcal{R}_0(L+1)=0$ to give
\begin{eqnarray}
\mathcal{R}_0(x)&=&a\left[(\lambda_+^x-1)-(\lambda_+^{L+1}-1)\frac{x}{L+1}\right]
+b\left[(\lambda_-^x-1)-(\lambda_-^{L+1}-1)\frac{x}{L+1}\right] ~,\nonumber\\
&\equiv&a\beta(x)+b\beta^*(x),
\end{eqnarray}
with $\beta(L+1)=0$.  The boundary condition $\mathcal{R}_0(L+2)=0$ leads to
$\mathcal{R}_0(x)=aV(x,L+2)$, where the Wronskian is now defined as
$V(x,y)=\beta(x)\beta^*(y)-\beta^*(x)\beta(y)$.  Imposing the boundary
condition $\mathcal{R}_0(L)=1$ gives the final result
\begin{equation}
\mathcal{R}_0(x)=\frac{V(x,L+2)}{V(L,L+2)}~.
\end{equation}
For the other two restricted exit probabilities, the same calculation as that
outlined above gives
\begin{equation}
\mathcal{R}_1(x)=\frac{W(x,L+2)}{W(L+1,L+2)}~,\qquad\qquad
\mathcal{R}_2(x)=\frac{W(x,L+1)}{W(L+2,L+1)}~.
\end{equation}
These results for the total and restricted exit probabilities are plotted in
Fig.~\ref{b23}.

\section{Exit probabilities for burst length $L/3<b<L/2$}
\label{app:longb}

When the burst length $b$ is in the range $[\frac{L}{3},\frac{L}{2}]$, the
interval naturally divides into the three subintervals $[0,L-2b-1]$,
$[L-2b,L-b-1]$, and $[L-b,L]$.  The recursion relations satisfied by the
total exit probability to the right edge of the interval are:
\begin{align}
\begin{split}
\label{3-int}
\mathcal{E}^{\rm I}(x)&=q\mathcal{E}^{\rm I}(x-1)+p\mathcal{E}^{\rm II}(x+b)\,,\\
\mathcal{E}^{\rm II}(x)&=q\mathcal{E}^{\rm II}(x-1)+p\mathcal{E}^{\rm III}(x+b)\,,\\
\mathcal{E}^{\rm III}(x)&=q\mathcal{E}^{\rm III}(x-1)+p\,.
\end{split}
\end{align}
These exit probabilities must also satisfy the joining and boundary
conditions
\begin{eqnarray*}
  \mathcal{E}^{\rm I}(0)&=&0\,,\nonumber \\
  \mathcal{E}^{\rm II}(L-2b)&=&q\mathcal{E}^{\rm I}(L-2b-1)+p\mathcal{E}^{\rm III}(L-b)\,,\nonumber\\
  \mathcal{E}^{\rm III}(L-b)&=&q\mathcal{E}^{\rm II}(L-b-1)+p\,.\nonumber
\end{eqnarray*}

We generalize the approach used to solve the two-interval case (cf.\
Eq.~\eqref{E2-result}) by first solving for $\mathcal{E}^{\rm III}$ in the
form $\mathcal{E}^{\rm III}=1+Aq^x$, substituting this result into the
recursion for $\mathcal{E}^{\rm II}$ to obtain its general form, and finally
substituting the result for $\mathcal{E}^{\rm II}$ into the recursion for
$\mathcal{E}^{\rm I}$.  All the unknown constants may then be fixed by the
boundary and joining conditions, and the final result is:
\begin{align}
\begin{split}
\mathcal{E}^{\rm I}(x)&=1-\frac{q^x\big\{pq^b \big[2(x-y)+pq^b (b+x-y) (b+x-y+1)\big]+2\big\}}{pq^b\big[(b-y)(b-y+1)
   pq^b-2y\big]+2}~,\\
\mathcal{E}^{\rm II}(x)&=1-\frac{2q^x \big[pq^b (x-y)+1\big]}{pq^b \big[(b-y) (b-y+1) pq^b-2y\big]+2}~,\\
\mathcal{E}^{\rm III}(x)&=1-\frac{2q^x}{pq^b\big[(b-y)(b-y+1)pq^b-2y\big]+2}~,
\end{split}
\end{align}
where $y= L-b-1$.  This procedure can be continued to as many subintervals as
desired both for the total and for the restricted exit probabilities.

\end{widetext}


\begin{thebibliography}{99}

\bibitem{nowak}M. Nowak and R. May, {\it Virus dynamics: mathematical
    principles of immunology and virology}, (Oxford University Press, New
  York, 2000).

\bibitem{perelson} A. S. Perelson, 
  Nat.\ Rev.\ Immunol.\ {\bf 2}, 28 (2002).

\bibitem{vk} N.~G.~Van~Kampen, {\it Stochastic Processes in Physics and
    Chemistry} (North-Holland, Amsterdam, 2001).

\bibitem{fpp} S. Redner, {\it A Guide to First-Passage Processes}, (Cambridge
University Press, New York, 2001).

\bibitem{PKP10} J. E. Pearson, P. L. Krapivsky, and A. S. Perelson, preprint.

\bibitem{AR06} T. Antal and S. Redner, J. Stat.\ Phys.\ {\bf 123}, 1129
  (2006).

\bibitem{left} For the left boundary, exit occurs only at $x=0$ and there is
  no distinction between the total and restricted exit probabilities.

\bibitem{GS01} G. Grimmett and D. Stirzaker, {\it Probability and Random
    Processes} 3rd ed., (Oxford University Press, Oxford, 2001).

\bibitem{DF87} B. Derrida and H. Flyvbjerg, J. Phys.\ A: Math.\ Gen.\ {\bf
    20}, 5273 (1987).

  \bibitem{FIK95} L. Frachebourg, I. Ispolatov, and P. L. Krapivsky, Phys.\
Rev.\ E {\bf 52}, R5727 (1995).

\end{thebibliography}
\end{document}